\documentclass[prb,aps, twocolumn, showpacs]{revtex4}
\usepackage{amssymb}
\usepackage{amsmath}
\usepackage[dvips]{graphicx}
\usepackage[english]{babel}

\begin{document}

\title{From one to $N$ Cooper pairs, step by step}
\author{W. V. Pogosov$^{1,2}$ and M. Combescot$^{1}$}
\affiliation{(1) Institut des NanoSciences de Paris, Universite Pierre et Marie Curie,
CNRS, Campus Boucicaut, 140 rue de Lourmel, 75015 Paris}
\affiliation{(2) Institute for Theoretical and Applied Electrodynamics, Russian Academy
of Sciences, Izhorskaya 13, 125412 Moscow}

\begin{abstract}
We extend the one-pair Cooper configuration towards
Bardeen-Cooper-Schrieffer (BCS) model of superconductivity by adding
one-by-one electron pairs to an energy layer where a small attraction acts.
To do it, we solve Richardson's equations analytically in the dilute limit
of pairs on the one-Cooper pair scale. We find, through only keeping the
first order term in this expansion, that the $N$ correlated pair energy
reads as the energy of $N$ isolated pairs within a $N(N-1)$ correction
induced by the Pauli exclusion principle which tends to decrease the average
pair binding energy when the pair number increases. Quite remarkably,
extension of this first-order result to the dense regime gives the BCS
condensation energy exactly. This leads us to suggest a different
understanding of the BCS condensation energy with a pair number equal to the
number of pairs feeling the potential and an average pair binding energy
reduced by Pauli blocking to half the single Cooper pair energy - instead of
the more standard but far larger superconducting gap.
\end{abstract}

\author{}
\maketitle
\date{\today }

\section{Introduction}

The possibility for an electron pair with up and down spins to form a bound
state under a very small attractive potential has been pointed out more than
50 years ago, first by Fr\"{o}hlich \cite{Frol} and then revealed by Cooper
\cite{Cooper} who considered adding a pair of electrons with opposite spins
and zero total momentum, to a full Fermi sea in an energy layer where a weak
attraction acts (see Fig. 1(a)). This attraction leads to the bound state
for the two added electrons known as a "Cooper pair". In a milestone paper
of the following year, Bardeen, Cooper and Schrieffer (BCS) \cite{BCS} have
succeeded to extend this idea to $N$ pairs. This was a crucial step toward
the understanding of superconductivity. The configuration they considered is
shown in Fig. 1(b): the attractive potential now acts in a layer extending
on both sides of thenoninteracting electrons Fermi level and pairs are
formed from electrons located in this layer. This BCS configuration can be
formally achieved by extending Cooper's model, through adding more and more
pairs to the "frozen" Fermi sea until this layer becomes half-filled. The
major difficulty to go from $1$ to $N$ pairs, step by step, comes from the
proper handling of the Pauli exclusion principle between a fixed number of
indistinguishable electrons. A smart way to overcome this difficulty is to
turn from the canonical ensemble in which $N$ is fixed, to the grand
canonical ensemble in which the mean value of $N$ is adjusted in the end to
fit the particle number. These two procedures are known to exactly give the
same results in the large $N$ limit - which is the physically relevant limit
for the problem at hand. For a discussion of the applicability of the
canonical ensemble to superconductivity, see, e.g., Refs. \cite%
{Bogoliubov,Delft}.

\begin{figure}[tbp]
\begin{center}
\includegraphics[width=0.4\textwidth]{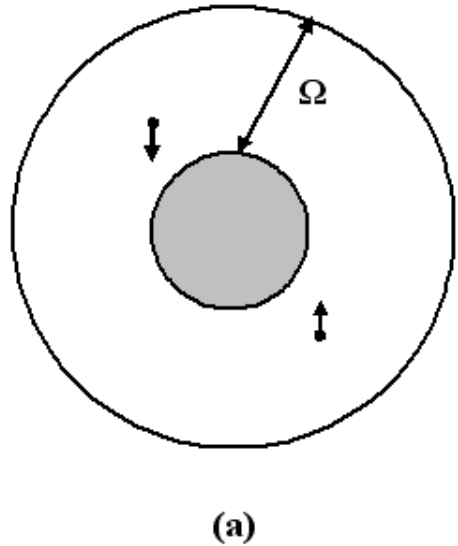}
\end{center}
\caption{Fig. 1(a). Frozen Fermi sea $\left\vert F_{0}\right\rangle $ (grey area) in
the one-Cooper pair problem and also in the problem we here address, the
number $N$ of pairs added to $\left\vert F_{0}\right\rangle $\ increasing,
one by one. A small attractive potential extends over an energy layer $%
\Omega $ above the Fermi sea $\left\vert F_{0}\right\rangle $. The two dots
schematically show the two electrons with opposite spins and momenta, of the
one-Cooper pair problem.}
\label{Fig1a}
\end{figure}

\begin{figure}[tbp]
\begin{center}
\includegraphics[width=0.4\textwidth]{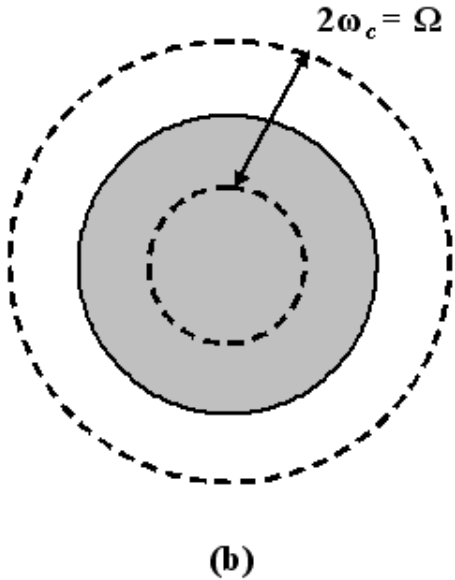}
\end{center}
\caption{ Fig. 1(b). Normal Fermi sea $\left\vert F\right\rangle $ (grey area) in the
usual BCS model. The potential extends symmetrically ($+\omega _{c}$, $%
-\omega _{c}$) on both sides of this Fermi sea, so that its total extension
is $2\omega _{c}=\Omega $. The Fermi sea $\left\vert F\right\rangle $ of
Fig. 1(b) contains $N=\rho _{0}\omega _{c}=\rho _{0}\Omega /2$ more
electrons than the frozen Fermi sea $\left\vert F_{0}\right\rangle $ which
is here shown as a dashed circle, the density of states $\rho _{0}$ being
taken as constant over $\Omega $.}
\label{Fig1b}
\end{figure}

A few years later, R.W. Richardson \cite{Rich1, Rich2, Rich3} has proposed a
way to stay in the canonical ensemble\cite{Bogoliubov1}. Through a elegant
procedure, he succeeded to analytically transform the Schr\"{o}dinger
equation for $N$ pairs into a coupled set of $N$ algebraic equations. These
equations, which are commonly called Richardson's equations (or more
properly Richardson-Gaudin's \cite{Gaudin} equations), have been solved
analytically in a few cases only: In Ref. \cite{Rich2}, R. W. Richardson
solved them for $N$ pairs in one and two multiply-degenerate levels. In
another work \cite{Rich3}, he used this procedure in the limit of an
infinite number of pairs within the half-filled configuration and succeeded
to rederive the BCS grand canonical result at the leading order in the
inverse particle number. Nowadays, Richardson's approach is widely used \cite%
{review} for the analysis of the superconducting properties of small
granules with nanometer sizes. Such granules accommodate countable number of
correlated electron pairs for which Richardson's equations can be solved
numerically. Conceptual difficulties in solving Richardson's equations can
be attributed to the fact that this procedure essentially corresponds to
reduce the quantum many-body problem to a classical one for "particles"
interacting via Coulomb-like forces. It however is well-known that the
analytical treatment of a many-body problem, even with classical particles,
stays difficult. In addition, a severe singularity problem arises when
solving these equations numerically \cite{review}, due to possible
cancelations in the denominators. This is why the resolution of Richardson's
equations in the general case, stayed an open problem up to now.

The purpose of the present work is to extend the original Cooper's work \cite%
{Cooper} from 1 to $N$ pairs within the canonical ensemble by solving
Richardson's equations analytically. We here calculate the ground state
energy of $N$ pairs when $N$ increases step by step starting from $N=1$. We
have recently derived the energy of just two pairs by various procedures
\cite{pap3} to point out the trend induced by Pauli blocking on Cooper pairs
in the simplest configurations. We have also given a sketch of the
derivation for general $N$ in Ref. \cite{pap1} where the analytical
expression of the energy of $N$ Cooper pairs was for the first time
reported. Although, at the present time, it seems difficult to perform such
a pair increase experimentally, this study can at least be seen as a
gedanken experiment to reveal a possible connection between two famous
problems, namely, Cooper's model and the BCS model for superconductivity, in
order to more deeply understand the role of the Pauli exclusion principle in
Cooper-paired states.

From a careful analysis of the structure of Richardson's equations for a
number of pairs going from 2 to 7, we have succeeded to see how these
equations can be manipulated to get the $N$-pair energy $E_{N}$ analytically
for $N=2n$ and $2n+1$, the structure of the solution depending on the parity
of $N$. The analytical solution we propose relies on an expansion in a
dimensionless parameter proportional to the inverse of the sample volume,
through a procedure similar to the one we have recently used to get the
energy of two pairs\cite{pap3}. This parameter actually is the inverse of
the number of pairs $N_{c}$ from which single Cooper pairs would start to
overlap. This limitates the validity of the expansion we perform to the
dilute regime, i.e., $N$ small compared to $N_{c}$ since we moreover only
keep the lowest order term in this expansion. This restriction a priori
prevents the extension of our result to the dense BCS regime in which half
the potential layer is filled by electrons. It however lets $N$ to be
arbitrary large in the thermodynamical limit because only $N/N_{c}$ matters.
In spite of this limitation, it is quite remarkable to note that the
analytical result we obtain, extended outside its mathematical validity
range, produces the BCS condensation energy \textit{exactly}, i.e., with all
its numerical prefactors, even when the potential does not extend
symmetrically on both sides of the Fermi level. This leads us to think that
our solution, although derived in the dilute limit, can be considered as an
important step towards mastering Richardson's approach in the general
context of superconductivity.

The expression we find for the ground state energy of $N$ pairs casts as%
\begin{equation}
E_{N}=NE_{1}+\frac{N(N-1)}{2}E_{int}  \label{1}
\end{equation}%
$E_{1}$ is the energy of one isolated pair, as originally found by Cooper,
while $E_{int}$\ can be seen as the interaction energy between two pairs. We
wish to stress that this interaction energy can only come from the Pauli
exclusion principle between the electrons from which these pairs are
constructed, because of the very peculiar form for the interaction potential
taken in the BCS theory of superconductivity which prevents any direct
interaction between pairs. Such a $N(N-1)$ term may appear as natural in the
dilute limit. However, no other terms in $N(N-1)(N-2)$ and higher ..., seem
to exist in the dense BCS regime, as the agreement of our dilute limit
result with the dense BCS one tends to indicate. This agreement, highly
unexpected at first, brings some new light to the concept of "Cooper pair".
In particular, it leads us to reconsider the standard understanding of the
BCS result for the ground state energy of the superconducting state in terms
of the excitation gap.

(i) The BCS result for the condensation energy (see, e.g., Ref. \cite%
{Fetter,Tinkham,Leggett}) is usually understood through the concept of
"virtual pairs", as introduced by Schrieffer. These pairs are made of
electrons with opposite spins and zero total momentum, excited above the
Fermi level of noninteracting "normal" electrons. From the energy scale $%
\Delta $ over which the pair density changes in the vicinity of this Fermi
level, one deduces a "virtual pair" number of the order of the number of
pairs within a thin layer of width $\Delta $. Such a number is far smaller
than the total number $N$ of electrons feeling the attractive potential and
used in the BCS wave function ansatz. In order to recover the BCS expression
of the ground state energy with such a small "virtual pair" number, it is
necessary to give to each of these pairs an energy of the order of $\Delta $%
. It however is clear that this "virtual pair" picture is highly
qualitative: these pairs are \textit{not} introduced ab initio in the BCS
theory and their number is not rigorously controlled, so that the binding
energy of a virtual pair is not a precisely defined quantity. These virtual
pairs essentially are a qualitative concept which provides a physical
picture for the BCS ground state energy \cite{BCS,Fetter,Tinkham}. It
actually is possible to introduce these excited pairs in a well defined
manner, as done by Leggett\cite{Leggett} using destruction and creation
operators acting not on the frozen Fermi sea, but on the normal electron
Fermi sea. However, this procedure does not yet generate a well defined pair
binding energy.

(ii) Our alternative understanding of the BCS result for the superconducting
ground state energy is based on the fact that correlations exist between
\textit{all} electrons which feel the attraction, so that all these
electrons must form pairs - not just a fraction as with the virtual pair
understanding described above. These pairs actually correspond to what
Schrieffer\cite{Schrieffer} calls "superfluid pairs". The pair number $N$
being then much larger than within the usual BCS understanding, the average
binding energy of these pairs ends by being much smaller than the energy gap
$\Delta $. We here show that it only is a fraction of the energy of one
\textit{isolated} Cooper pair. More precisely, this average binding energy
decreases linearly when the number of pairs increases. This decrease
entirely comes from the Pauli exclusion principle that restricts the number
of available excited states when constructing paired states: with less and
less empty states enjoying the attractive potential, the correlated pairs
gain a smaller and smaller average binding energy, as if each of the empty
states were bringing a constant elementary contribution to this binding
energy. Our present work actually shows that, in spite of the widely spread
idea that the isolated pair picture has little meaning in the dense regime%
\cite{Schrieffer}, it is possible to recover the BCS ground state energy in
an exact way, by simply using the concept of isolated Cooper pair but by
properly including Pauli blocking between the electrons of these pairs.

We also would like to mention that the result for the ground state energy we
found, is in agreement with the solution of Richardson's equations in the
large-sample limit obtained earlier \cite{Rich2}, except that we here
consider arbitrary fillings of the potential layer, in contrast to Ref. \cite%
{Rich2} which focuses on the half-filling configuration only. Changing the
layer filling enables us to explicitly establish the link between the Cooper
model and the BCS picture, as they are presented in textbooks. Moreover, our
procedure allows us to reveal an important underlying physics which is
somehow masked when considering only one particular configuration, namely,
half filling this configuration being, actually, quite specific. In
addition, the analytical approach of Ref. \cite{Rich2}, as well as more
recent studies\cite{Roman,Altshuler} based on the same approach, assumes
that Richardson's energy-like quantities in the ground state are organized into an arc in a
complex plane. This very strong assumption is based on some numerics on
Richardson's equations\cite{Rich2}. However, in our opinion, this assumption
is not fully controllable, so that it is not guaranteed that it does not
break in certain region of the phase diagram (for instance, in the delicate
BEC-BCS crossover region). Therefore, alternative methods for solving
Richardson equations, as the one we here proposed, are highly desirable.

The present paper is organized as follows.

In Section II, we settle the problem. We outline Richardson's procedure to
transform the Schr\"{o}dinger equation for 2 Cooper pairs into two nonlinear
equations and we report how this procedure extends to $N$ pairs. We then
explain how an analytical solution of these Richardson's equations can be
obtained, thanks to the identification of a small dimensionless parameter.
In practice, this is done through the expansion of sums similar to the one
which appears in the one-Cooper pair problem.

In Section III, we use this expansion to explicitly solve Richardson's
equations analytically, step by step, from 2 to 7 pairs. We determine the $N$
energy-like quantities which are solution of these equations, at the lowest
order in the dimensionless parameter we have previously identified. A
careful analysis of these solutions allows us to construct a general form
for the energy of an arbitrary number of pairs. The structure of the
equations for $2n$ and $2n+1$ being different, the solution for 6 and 7
pairs corresponds to $n=3$ which is somewhat generic. In contrast, $n=2$ for
4 and 5 pairs is not because $n$ and $n!$ are then equal. This explains why
we have had to go up to (6, 7) to fully control how the general solution
develops. In this section, we also discuss the validity of the procedure we
propose to solve these equations.

In Section IV, we reconsider the condensation energy obtained by Bardeen,
Cooper, Schrieffer and its usual understanding in terms of "Cooper pairs",
in the light of the result we have obtained through the Richardson's
canonical ensemble procedure.

We conclude in Section V.

\section{Exact form of the $N$-pair energy}

\subsection{The problem}

We consider a "frozen Fermi sea" $\left\vert F_{0}\right\rangle $\ made of $%
N_{F_{0}}$\ free electron pairs with up and down spins, their energy
extending from zero to $\varepsilon _{F_{0}}=k_{_{F_{0}}}^{2}/2m$, and we
start adding zo-momentum electron pairs to this Fermi sea in order to extend
Cooper's model towards the traditional BCS configuration.

Electrons outside $\left\vert F_{0}\right\rangle $ feel an attractive
potential $\mathcal{V}$ that we take as being the standard potential used in
the BCS theory of superconductivity, without questioning its
oversimplifications since we want to position the present work within the
very first version of BCS theory. This potential reads as

\begin{eqnarray}
\mathcal{V} &=&\sum_{\mathbf{k},\mathbf{k}^{\prime }}V_{\mathbf{k}^{\prime }%
\mathbf{k}}a_{\mathbf{k}^{\prime }\uparrow }^{\dagger }a_{-\mathbf{k}%
^{\prime }\downarrow }^{\dagger }a_{-\mathbf{k}\downarrow }a_{\mathbf{k}%
\uparrow }  \notag \\
&=&\sum_{\mathbf{k},\mathbf{k}^{\prime }}V_{\mathbf{k}^{\prime }\mathbf{k}%
}\beta _{\mathbf{k}^{\prime }}^{\dagger }\beta _{\mathbf{k}}  \label{2}
\end{eqnarray}%
$V_{\mathbf{k}^{\prime }\mathbf{k}}$ is further on approximated by a contact
separable potential%
\begin{equation}
V_{\mathbf{k}^{\prime }\mathbf{k}}=-Vw_{\mathbf{k}^{\prime }}w_{\mathbf{k}}
\label{3}
\end{equation}%
with $V$ being a positive constant while $w_{\mathbf{k}}=1$\ inside the
potential layer, i.e., for $\varepsilon _{F_{0}}<\varepsilon _{\mathbf{k}%
}<\varepsilon _{F_{0}}+\Omega $, and 0 otherwise. Thus, this potential is
only felt by electrons with opposite spins and opposite momenta, outside the
Fermi sea $\left\vert F_{0}\right\rangle $, up to a cut off that we have
chosen to call $\varepsilon _{F_{0}}+\Omega $. The system Hamiltonian then
reads as $H=H_{0}+\mathcal{V}$ where
\begin{equation}
H_{0}=\sum_{\mathbf{k}}\varepsilon _{\mathbf{k}}\left( a_{\mathbf{k}\uparrow
}^{\dagger }a_{\mathbf{k}\uparrow }+a_{\mathbf{k}\downarrow }^{\dagger }a_{%
\mathbf{k}\downarrow }\right)   \label{4}
\end{equation}

We add $N=1,$ $2,$ $3$... pairs of electrons with up and down spins and zero
total momentum to this Fermi sea $\left\vert F_{0}\right\rangle $ and we
look for the exact ground state energy of these $N$ pairs resulting from
both the potential $\mathcal{V}$ and the Pauli exclusion principle.

\subsection{One Cooper pair}

The $N=1$ case is the standard problem studied by Cooper. By writing the
wave function as
\begin{equation}
\left\vert \psi _{1}\right\rangle =\sum w_{\mathbf{k}}G(\mathbf{k})a_{%
\mathbf{k}\uparrow }^{\dagger }a_{-\mathbf{k}\downarrow }^{\dagger
}\left\vert F_{0}\right\rangle  \label{5}
\end{equation}%
where $\left\vert F_{0}\right\rangle $\ is the previously defined "frozen
Fermi sea", we find that $\left\vert \psi _{1}\right\rangle $\ fulfills the
Schr\"{o}dinger equation $\left( H-E_{1}\right) \left\vert \psi
_{1}\right\rangle =0$, provided that the $G(\mathbf{k})$'s are such that%
\begin{equation}
G(\mathbf{k})\left( 2\varepsilon _{\mathbf{k}}-E_{1}\right) -V\sum_{\mathbf{p%
}}w_{\mathbf{p}}G(\mathbf{p})=0  \label{6}
\end{equation}%
since $w_{\mathbf{k}}=w_{\mathbf{k}}^{2}$. If we now extract $G(\mathbf{k})$
from the above equation, multiply it by $w_{\mathbf{k}}$ and sum over $%
\mathbf{k}$, we end with the textbook equation for the one-Cooper pair
energy, namely%
\begin{equation}
1=V\sum_{\mathbf{k}}\frac{w_{\mathbf{k}}}{2\varepsilon _{\mathbf{k}}-E_{1}}
\label{7}
\end{equation}

To calculate this one-pair energy explicitly, we turn to the continuous
limit. If the density of states $\rho (\varepsilon )$\ does not change too
much over the potential extension $\varepsilon _{F_{0}}<\varepsilon _{%
\mathbf{k}}<$\ $\varepsilon _{F_{0}}+\Omega $, we get
\begin{eqnarray}
V\sum_{\mathbf{k}}\frac{w_{\mathbf{k}}}{2\varepsilon _{\mathbf{k}}-E}
&\simeq &\frac{\rho _{0}V}{2}\int_{\varepsilon _{F_{0}}}^{\varepsilon
_{F_{0}}+\Omega }\frac{2d\varepsilon }{2\varepsilon -E}  \notag \\
&=&(v/2)\ln \left\vert \frac{2\varepsilon _{F_{0}}+2\Omega -E}{2\varepsilon
_{F_{0}}-E}\right\vert   \label{8}
\end{eqnarray}%
where $\rho _{0}$ is the average density of states in the potential layer,
while $v=\rho _{0}V$ is the small dimensionless parameter which
characterizes the weak potential felt by electrons above $\varepsilon
_{F_{0}}$.When inserted into Eq. (7), this gives the binding energy of one
"Cooper pair" as%
\begin{eqnarray}
E_{1} &=&2\varepsilon _{F_{0}}-\epsilon _{c}  \notag \\
\epsilon _{c} &=&2\Omega y_{c}  \label{9}
\end{eqnarray}%
where $y_{c}$ is an extremely small positive constant related to $v$\ through%
\begin{equation}
y_{c}=e^{-2/v}\left( 1-e^{-2/v}\right) ^{-1}\simeq e^{-2/v}  \label{10}
\end{equation}%
in the weak coupling limit, i.e., $v\ll 1$. This shows that the binding
energy of one Cooper pair depends in a singular nonpertubative way on the
small attractive potential $V$. This also shows, more importantly for the
problem we here address, that this binding energy depends linearly on the
potential extension $\Omega $, i.e., on the number of empty electron states
feeling the attractive potential. To evidence this understanding, we can
rewrite $\epsilon _{c}$ as%
\begin{equation}
\epsilon _{c}=\left( \rho _{0}\Omega \right) \left( \frac{\epsilon _{c}}{%
\rho _{0}\Omega }\right) =N_{\Omega }\epsilon _{V}  \label{11}
\end{equation}%
where $N_{\Omega }=\rho _{0}\Omega $ is the number of empty pair states for
electrons with up or down spin in the energy layer $\Omega $ while $\epsilon
_{V}$ given by%
\begin{equation}
\epsilon _{V}\simeq \frac{2}{\rho _{0}}e^{-2/\rho _{0}V}  \label{12}
\end{equation}%
can be understood as the contribution of each of these empty pair states to
the pair binding energy. This remark is going to be quite important for the
understanding of what really happens when the number of added pairs
increases to possibly fill half the potential layer - as in the usual BCS
configuration - or even the whole layer, which is an interesting
mathematical limit. Actually, a binding energy proportional to the number of
empty states $N_{\Omega }$, as in Eq. (11), already makes us guessing that
the average binding energy in the $N$-pair configuration should decrease
linearly when the number of pairs increases, due to the decrease of empty
states available in the linear combination responsible for the single-pair
binding energy. With this in mind, we now study 2 pairs, then 3 pairs and so
on, in order to understand how the energy changes with $N$.

\subsection{Two Cooper pairs and the Richardson's procedure}

Let us first rederive Richardson's equations for the reader not fully aware
of them since the present work heavily relies on these equations. The
derivation we here propose is somewhat different from the original one.

The very peculiar form of the potential $\mathcal{V}$ given in Eq. (2) leads
us to look for the ground state of two pairs as%
\begin{equation}
\left\vert \psi _{2}\right\rangle =\sum w_{\mathbf{k}_{1}}w_{\mathbf{k}%
_{2}}G(\mathbf{k}_{1},\mathbf{k}_{2})\beta _{\mathbf{k}_{1}}^{\dagger }\beta
_{\mathbf{k}_{2}}^{\dagger }\left\vert F_{0}\right\rangle  \label{13}
\end{equation}%
with $G(\mathbf{k}_{1},\mathbf{k}_{2})=G(\mathbf{k}_{2},\mathbf{k}_{1})$, as
always possible to enforce due to bosonic symmetry in fermion pair exchange,
namely, $\beta _{\mathbf{k}_{1}}^{\dagger }\beta _{\mathbf{k}_{2}}^{\dagger
}=\beta _{\mathbf{k}_{2}}^{\dagger }\beta _{\mathbf{k}_{1}}^{\dagger }$.

We then note that for two free pairs, we have
\begin{equation}
H_{0}\beta _{\mathbf{k}_{1}}^{\dagger }\beta _{\mathbf{k}_{2}}^{\dagger
}\left\vert F_{0}\right\rangle =\left( 2\varepsilon _{\mathbf{k}%
_{1}}+2\varepsilon _{\mathbf{k}_{2}}\right) \beta _{\mathbf{k}_{1}}^{\dagger
}\beta _{\mathbf{k}_{2}}^{\dagger }\left\vert F_{0}\right\rangle   \label{14}
\end{equation}%
\begin{equation}
\mathcal{V}\beta _{\mathbf{k}_{1}}^{\dagger }\beta _{\mathbf{k}%
_{2}}^{\dagger }\left\vert F_{0}\right\rangle =  \notag
\end{equation}%
\begin{equation}
-V(1-\delta _{\mathbf{k}_{1}\mathbf{k}_{2}})\left( w_{\mathbf{k}_{2}}\beta _{%
\mathbf{k}_{1}}^{\dagger }+w_{\mathbf{k}_{1}}\beta _{\mathbf{k}%
_{2}}^{\dagger }\right) \sum_{\mathbf{p}}w_{\mathbf{p}}\beta _{\mathbf{p}%
}^{\dagger }\left\vert F_{0}\right\rangle   \label{15}
\end{equation}%
the prefactor $(1-\delta _{\mathbf{k}_{1}\mathbf{k}_{2}})$ in the RHS of the
above equation being necessary to enforce the cancellation of this RHS for $%
\mathbf{k}_{1}=\mathbf{k}_{2}$, the state $\beta _{\mathbf{k}_{1}}^{\dagger
2}\left\vert F_{0}\right\rangle $ then reducing to zero. As $G(\mathbf{k}%
_{1},\mathbf{k}_{2})=G(\mathbf{k}_{2},\mathbf{k}_{1})$, this leads to%
\begin{equation}
0=\left( H-E_{2}\right) \left\vert \psi _{2}\right\rangle =\sum F(\mathbf{k}%
_{1},\mathbf{k}_{2})\beta _{\mathbf{k}_{1}}^{\dagger }\beta _{\mathbf{k}%
_{2}}^{\dagger }\left\vert F_{0}\right\rangle   \label{16}
\end{equation}%
where we have set
\begin{equation}
F(\mathbf{k}_{1},\mathbf{k}_{2})=w_{\mathbf{k}_{1}}w_{\mathbf{k}_{2}}[\left(
2\varepsilon _{\mathbf{k}_{1}}+2\varepsilon _{\mathbf{k}_{2}}-E_{2}\right) G(%
\mathbf{k}_{1},\mathbf{k}_{2})  \notag  \label{17}
\end{equation}%
\begin{equation}
-V\sum_{\mathbf{k}\neq \mathbf{k}_{1}}w_{\mathbf{k}}G(\mathbf{k},\mathbf{k}%
_{1})-V\sum_{\mathbf{k}\neq \mathbf{k}_{2}}w_{\mathbf{k}}G(\mathbf{k},%
\mathbf{k}_{2})]
\end{equation}%
By using the scalar product of two-free pair states
\begin{eqnarray*}
\left\langle F_{0}\right\vert \beta _{\mathbf{p}_{2}}\beta _{\mathbf{p}%
_{1}}\beta _{\mathbf{k}_{1}}^{\dagger }\beta _{\mathbf{k}_{2}}^{\dagger
}\left\vert F_{0}\right\rangle  &=&\left( 1-\delta _{\mathbf{p}_{1}\mathbf{p}%
_{2}}\right)  \\
&&\left( \delta _{\mathbf{p}_{1}\mathbf{k}_{1}}\delta _{\mathbf{p}_{2}%
\mathbf{k}_{2}}+\delta _{\mathbf{p}_{1}\mathbf{k}_{2}}\delta _{\mathbf{p}_{2}%
\mathbf{k}_{1}}\right)
\end{eqnarray*}%
we find that $\left\vert \psi _{2}\right\rangle $ is solution of the Schr%
\"{o}dinger equation provided that
\begin{equation}
\left( 1-\delta _{\mathbf{p}_{1}\mathbf{p}_{2}}\right) F(\mathbf{p}_{1},%
\mathbf{p}_{2})=0  \label{18}
\end{equation}%
This makes $G(\mathbf{p}_{1},\mathbf{p}_{2})$ undefined for $\mathbf{p}_{1}=%
\mathbf{p}_{2}$, which is unimportant since the state $\beta _{\mathbf{k}%
_{1}}^{\dagger }\beta _{\mathbf{k}_{2}}^{\dagger }\left\vert
F_{0}\right\rangle $ reduces to zero for $\mathbf{k}_{1}=\mathbf{k}_{2}$
anyway. In contrast, $G(\mathbf{k}_{1},\mathbf{k}_{2})$ for $\mathbf{k}%
_{1}\neq \mathbf{k}_{2}$ must fulfill the integral equation $F(\mathbf{k}%
_{1},\mathbf{k}_{2})=0$. The really unpleasant part of this integral
equation is the coupling between ($\mathbf{k}_{1}$, $\mathbf{k}_{2}$)
through $\left( 2\varepsilon _{\mathbf{k}_{1}}+2\varepsilon _{\mathbf{k}%
_{2}}-E_{2}\right) $. To decouple $\mathbf{k}_{1}$ from $\mathbf{k}_{2}$,
Richardson suggested splits $E_{2}$ as $R_{1}+R_{2}$. In view of the
structure of the equation $F(\mathbf{k}_{1},\mathbf{k}_{2})=0$, we are led
to find a symmetrical function of ($\mathbf{k}_{1}$, $\mathbf{k}_{2}$)
which, multiplied by $\left( 2\varepsilon _{\mathbf{k}_{1}}+2\varepsilon _{%
\mathbf{k}_{2}}-E_{2}\right) $, splits into a $\mathbf{k}_{1}$ function plus
a $\mathbf{k}_{2}$ function. To do it, we first note that this unpleasant
factor also appears in the identity
\begin{equation}
\frac{1}{2\varepsilon _{\mathbf{k}_{1}}-R_{1}}+\frac{1}{2\varepsilon _{%
\mathbf{k}_{2}}-R_{2}}=\frac{2\varepsilon _{\mathbf{k}_{1}}+2\varepsilon _{%
\mathbf{k}_{2}}-E_{2}}{\left( 2\varepsilon _{\mathbf{k}_{1}}-R_{1}\right)
\left( 2\varepsilon _{\mathbf{k}_{2}}-R_{2}\right) }  \label{19}
\end{equation}%
and a similar identity with ($R_{1}$, $R_{2}$) exchanged. Since we look for $%
G(\mathbf{k}_{1},\mathbf{k}_{2})$ as a symmetrical function of ($\mathbf{k}%
_{1}$, $\mathbf{k}_{2}$), we are naturally led to consider%
\begin{equation}
G(\mathbf{k}_{1},\mathbf{k}_{2})=\frac{1}{\left( 2\varepsilon _{\mathbf{k}%
_{1}}-R_{1}\right) \left( 2\varepsilon _{\mathbf{k}_{2}}-R_{2}\right) }
\notag
\end{equation}%
\begin{equation}
+\frac{1}{\left( 2\varepsilon _{\mathbf{k}_{1}}-R_{2}\right) \left(
2\varepsilon _{\mathbf{k}_{2}}-R_{1}\right) }  \label{20}
\end{equation}%
Indeed, when multiplied by $\left( 2\varepsilon _{\mathbf{k}%
_{1}}+2\varepsilon _{\mathbf{k}_{2}}-E_{2}\right) $, the resulting quantity
splits into $\left[ \left( 2\varepsilon _{\mathbf{k}_{1}}-R_{1}\right)
^{-1}+\left( 2\varepsilon _{\mathbf{k}_{1}}-R_{2}\right) ^{-1}\right] $ plus
a similar term with $\mathbf{k}_{1}$\ replaced by $\mathbf{k}_{2}$. We then
find that this $G(\mathbf{k}_{1},\mathbf{k}_{2})$\ function fulfills Eq.
(16), provided that
\begin{equation}
w_{\mathbf{k}_{1}}\left( \frac{1}{2\varepsilon _{\mathbf{k}_{1}}-R_{1}}+%
\frac{1}{2\varepsilon _{\mathbf{k}_{1}}-R_{2}}\right) =Vw_{\mathbf{k}%
_{1}}\sum_{\mathbf{k}\neq \mathbf{k}_{1}}w_{\mathbf{k}}G(\mathbf{k},\mathbf{k%
}_{1})  \label{21}
\end{equation}%
for arbitrary $\mathbf{k}_{1}$. The next step is to get rid of $\mathbf{k}%
_{1}$. For that, we first insert Eq. (20) for $G$ into the above equation.
This leads to%
\begin{eqnarray}
&&\frac{w_{\mathbf{k}_{1}}}{2\varepsilon _{\mathbf{k}_{1}}-R_{1}}\left[
1-V\sum_{\mathbf{k}}\frac{w_{\mathbf{k}}}{2\varepsilon _{\mathbf{k}}-R_{2}}%
\right]   \notag \\
&&+\frac{w_{\mathbf{k}_{1}}}{2\varepsilon _{\mathbf{k}_{1}}-R_{2}}\left[
1-V\sum_{\mathbf{k}}\frac{w_{\mathbf{k}}}{2\varepsilon _{\mathbf{k}}-R_{1}}%
\right]   \notag \\
&=&-\frac{2w_{\mathbf{k}_{1}}}{\left( 2\varepsilon _{\mathbf{k}%
_{1}}-R_{1}\right) \left( 2\varepsilon _{\mathbf{k}_{1}}-R_{2}\right) }
\label{22}
\end{eqnarray}%
where the sums are now taken over all $\mathbf{k}$. By noting that the RHS
of the above equation also reads%
\begin{equation*}
-\frac{2w_{\mathbf{k}_{1}}}{R_{1}-R_{2}}\left( \frac{1}{2\varepsilon _{%
\mathbf{k}_{1}}-R_{1}}-\frac{1}{2\varepsilon _{\mathbf{k}_{1}}-R_{2}}\right)
\end{equation*}%
for $R_{1}\neq R_{2}$, as always possible to enforce since, up to now, the
only requirement is to have ($R_{1}$, $R_{2}$) such that $E_{2}=R_{1}+R_{2}$%
, we then find that Eq. (22) leads to%
\begin{equation}
\frac{f(R_{1},R_{2})}{2\varepsilon _{\mathbf{k}_{1}}-R_{1}}+\frac{%
f(R_{2},R_{1})}{2\varepsilon _{\mathbf{k}_{1}}-R_{2}}=0  \label{23}
\end{equation}%
where $f(R_{1},R_{2})$ is given by%
\begin{equation}
f(R_{1},R_{2})=1-V\sum_{\mathbf{k}}\frac{w_{\mathbf{k}}}{2\varepsilon _{%
\mathbf{k}}-R_{1}}-\frac{2V}{R_{1}-R_{2}}  \label{24}
\end{equation}%
Since $R_{1}\neq R_{2}$, the validity of Eq. (23) for arbitrary $\mathbf{k}%
_{1}$ imposes $f(R_{1},R_{2})=0=f(R_{2},R_{1})$. So that we end with
\begin{eqnarray}
1 &=&V\sum_{\mathbf{p}}\frac{w_{\mathbf{p}}}{2\varepsilon _{\mathbf{p}}-R_{1}%
}+\frac{2V}{R_{1}-R_{2}}  \notag \\
1 &=&V\sum_{\mathbf{p}}\frac{w_{\mathbf{p}}}{2\varepsilon _{\mathbf{p}}-R_{2}%
}+\frac{2V}{R_{2}-R_{1}}  \label{25}
\end{eqnarray}

These two equations are known as the Richardson's equations for two pairs.
They transform the resolution of the Schr\"{o}dinger equation for two pairs
into the resolution of two coupled algebraic equations for ($R_{1}$, $R_{2}$%
), a far simpler problem, definitly. We wish to stress that the existence of
an exact solution to the 2-pair Schr\"{o}dinger equation is deeply linked to
the very peculiar form of the reduced BCS interaction potential, the
zero-momentum pair $\beta _{\mathbf{k}}^{\dagger }$ staying unbroken by this
interaction.

\subsection{Richardson's equations for $N$ Cooper pairs}

We now consider the Fermi sea $\left\vert F_{0}\right\rangle $ plus three
electron pairs with up and down spins. A similar procedure leads us to split
the 3-pair energy as $E_{3}=R_{1}+R_{2}+R_{3}$, with $\left( R_{1},\text{ }%
R_{2},\text{ }R_{3}\right) $ fulfilling the three equations%
\begin{eqnarray}
1 &=&V\sum_{\mathbf{p}}\frac{w_{\mathbf{p}}}{2\varepsilon _{\mathbf{p}}-R_{1}%
}+\frac{2V}{R_{1}-R_{2}}+\frac{2V}{R_{1}-R_{3}}  \notag \\
1 &=&V\sum_{\mathbf{p}}\frac{w_{\mathbf{p}}}{2\varepsilon _{\mathbf{p}}-R_{2}%
}+\frac{2V}{R_{2}-R_{1}}+\frac{2V}{R_{2}-R_{3}}  \notag \\
1 &=&V\sum_{\mathbf{p}}\frac{w_{\mathbf{p}}}{2\varepsilon _{\mathbf{p}}-R_{3}%
}+\frac{2V}{R_{3}-R_{1}}+\frac{2V}{R_{3}-R_{2}}  \label{26}
\end{eqnarray}%
And so on for a higher number of pairs.

\subsection{Expansion of the relevant sum}

The unpleasant parts of these equations are the sums over ${\mathbf{p}}$. It
is in fact possible to tackle this set of nonlinear equations analytically
by expanding these sums close to the one-Cooper pair energy $E_{1}$ defined
in Eq. (7), since for $R=E_{1}$, the sum reduces to $1/V$. A convenient way
to do it, is to note that%
\begin{eqnarray}
&&V\sum_{\mathbf{p}}\frac{w_{\mathbf{p}}}{2\varepsilon _{\mathbf{p}}-R}
\notag \\
&=&V\sum_{\mathbf{p}}w_{\mathbf{p}}\left[ \frac{1}{2\varepsilon _{\mathbf{p}%
}-E_{1}}+\frac{R-E_{1}}{\left( 2\varepsilon _{\mathbf{p}}-E_{1}\right)
\left( 2\varepsilon _{\mathbf{p}}-R\right) }\right]   \notag \\
&=&1+\sum_{m=1}^{\infty }\sigma _{m}\left( R-E_{1}\right) ^{m}  \label{27}
\end{eqnarray}%
The prefactor $\sigma _{m}$ in this expansion reads in terms of the one-pair
energy $E_{1}$\ as%
\begin{equation}
\sigma _{m}=V\sum_{\mathbf{p}}\frac{w_{\mathbf{p}}}{\left( 2\varepsilon _{%
\mathbf{p}}-E_{1}\right) ^{m+1}}=\frac{v}{2m}\frac{I_{m}}{\epsilon _{c}^{m}}
\label{28}
\end{equation}%
where $\epsilon _{c}$ is the binding energy of one Cooper pair defined in
Eq. (11), while $I_{m}$ is found to read in terms of $y_{c}$\ defined in Eq.
(10) as%
\begin{equation}
I_{m}=1-\left( \frac{y_{c}}{1+y_{c}}\right) ^{m}=1-e^{-2m/v}  \label{29}
\end{equation}%
This makes it very close but not exactly equal to 1. Difference with 1 is
going to be crucial in the final expression of the $N$-pair energy. In view
of Eq. (28), we are led to rescale the energy difference in Eq. (27) as%
\begin{equation}
R-E_{1}=z\epsilon _{c}  \label{30}
\end{equation}%
This allows us to ultimately write the sums appearing in Richardson's
equations in a dimensionless form as%
\begin{equation}
V\sum_{\mathbf{p}}\frac{w_{\mathbf{p}}}{2\varepsilon _{\mathbf{p}}-R}%
=1+v\sum_{m=1}^{\infty }\frac{I_{m}}{m}z^{m}  \label{31}
\end{equation}%
this expansion being a priori valid for $\left\vert z\right\vert $\ small
compared to 1.

\section{Analytical solution of Richardson's equations}

We now turn to the resolution of Richardson's equations starting from two
pairs, as given in Eq. (25).

\subsection{Two Cooper pairs}

By using the expansion given in Eq. (31), we can transform the two
Richardson's equations for 2 pairs as%
\begin{eqnarray}
0 &=&\sum_{m=1}\frac{I_{m}}{m}z_{1}^{m}+\frac{\gamma }{z_{1}-z_{2}}
\label{32} \\
0 &=&\sum_{m=1}\frac{I_{m}}{m}z_{2}^{m}+\frac{\gamma }{z_{2}-z_{1}}
\label{33}
\end{eqnarray}%
where the $z_{i}$'s\ are related to the $R_{i}$'s through Eq. (30), namely, $%
R_{i}-E_{1}=z_{i}\epsilon _{c}$. The dimensionless parameter $\gamma $\ in
these equations appears to be%
\begin{equation}
\gamma =\frac{4}{\rho _{0}\epsilon _{c}}=\frac{4}{N_{c}}  \label{34}
\end{equation}%
where $N_{c}=\rho _{0}\epsilon _{c}$\ is the number of states within a
one-Cooper pair binding energy layer at the frozen Fermi sea level. Although
small compared to the number of pairs $N_{\Omega }$ in the energy layer $%
\Omega $ where the potential acts, $N_{c}$ is definitely very large compared
to 1. Indeed, in a sample volume $L^{D}$, the density of states $\rho _{0}$
at the Fermi level with momentum $k_{F}$ is of the order of $%
L^{D}mk_{F}^{D-2}$, where $D=3$ or 2 is the space dimension. This gives $%
N_{c}$ of the order of $L^{D}/l_{c}^{2}l_{F}^{D-2}$\ where $l_{F}=k_{F}^{-1}$
is the distance between Fermi sea electrons and $l_{c}$, defined as $%
\epsilon _{c}\sim 1/m_{c}l_{c}^{2}$, scales as the distance between the two
electrons forming one Cooper pair. This makes $N_{c}$ infinitely large
compared to 1 in the infinite-size sample limit. Consequently $\gamma $,
defined in Eq. (34), is a small parameter for macroscopic samples.

The existence of such a small dimensionless parameter is the key for solving
Richardson's equations analytically. Indeed, in order for the set of Eqs.
(32,33) to have a solution with $z_{i}$ small, as necessary for the
expansion (31) to be valid, we must have $z_{1}-z_{2}$ of the order of $%
\sqrt{\gamma }$. By replacing one of the two equations (32,33) by their sum,
namely
\begin{equation}
0=\sum_{m=1}\frac{I_{m}}{m}\left( z_{1}^{m}+z_{2}^{m}\right)   \label{35}
\end{equation}%
we find from the $m=1$ term of this equation that $z_{1}+z_{2}$\ must be
equal to zero at the $\sqrt{\gamma }$ order. This imposes to look for $z_{1}$%
\ and $z_{2}$\ as $z_{1}\simeq -z_{2}\simeq a_{1}\sqrt{\gamma }$. When
replaced into Eq. (32) in which we only keep the $m=1$ term, we find $%
0\simeq I_{1}a_{1}+1/2a_{1}$; so that $a_{1}^{2}\simeq -1/2I_{1}$. This
readily shows that $a_{1}$ is imaginary at the $\sqrt{\gamma }$ order.

To get a nonzero value for the sum $z_{1}+z_{2}$, as necessary to find the
energy change induced by interaction between Cooper pairs, i.e., to have $%
E_{2}$ not exactly equal to $2E_{1}$, we must keep two terms in Eq. (35).
This gives%
\begin{equation}
0\simeq I_{1}(z_{1}+z_{2})+\frac{I_{2}}{2}\left( z_{1}^{2}+z_{2}^{2}\right)
\label{36}
\end{equation}%
so that $(z_{1}+z_{2})$\ to lowest order in $\gamma $ reads as%
\begin{equation}
z_{1}+z_{2}\simeq -\frac{I_{2}}{2I_{1}}\left( z_{1}^{2}+z_{2}^{2}\right)
\simeq \frac{1}{2}\gamma \frac{I_{2}}{I_{1}^{2}}  \label{37}
\end{equation}%
If we now note that $(\epsilon _{c}\gamma )$ is just $(4/\rho _{0})$, as
seen from Eq. (34), the above equation gives the energy of two Cooper pairs $%
E_{2}=R_{1}+R_{2}$ as%
\begin{equation}
E_{2}\simeq 2\left( E_{1}+\frac{1}{\rho _{0}}\frac{I_{2}}{I_{1}^{2}}\right)
\label{38}
\end{equation}%
Since $I_{2}/I_{1}^{2}\simeq 1+2y_{c}$, while $2y_{c}\rho _{0}=\epsilon
_{c}/N_{\Omega }$, this energy appears in terms of physically relevant
quantities as%
\begin{eqnarray}
E_{2} &\simeq &2\left[ \left( 2\varepsilon _{F_{0}}+\frac{1}{\rho _{0}}%
\right) -\epsilon _{c}\left( 1-\frac{1}{N_{\Omega }}\right) \right]  \notag
\\
&\simeq &2\left[ \left( 2\varepsilon _{F_{0}}+\frac{1}{\rho _{0}}\right)
-2(N_{\Omega }-1)\epsilon _{V}\right]  \label{39}
\end{eqnarray}

As previously introduced, $N_{\Omega }=\rho _{0}\Omega $\ is the number of
pair states in the potential extension $\Omega $ above the Fermi sea $%
\left\vert F_{0}\right\rangle $, while $\epsilon _{V}$ is the elementary
contribution to the pair binding energy coming from each of the empty pair
states feeling the potential, their number being reduced to $(N_{\Omega }-1)$
in the 2-pair configuration, due to the added pair .

The above equations show that the energy change from the energy $2E_{1}$ of
two isolated Cooper pairs has a naive term $2/\rho _{0}$\ which simply comes
from the increase of the Fermi sea level when a second pair is added. It
also has a more subtle term $2\epsilon _{c}/N_{\Omega }=4\epsilon _{V}$\
which can be seen as twice the shrinkage of the one-Cooper pair binding
energy coming from the Pauli exclusion principle between electrons with up
spin and between electrons with down spin, available to form Cooper pairs
within the potential extension $\Omega $. Such a "moth-eaten effect" which
here exists for Cooper pairs, has already been found in the case of
composite bosons like Wannier or Frenkel excitons \cite{Monique}.

Let us now see how this energy change generalizes when the number of added
pairs increases from $N=2$. Since the analytical resolution of Richardson's
equations has raised problems for decades, we are going to consider $%
N=3,4,...$ pairs successively in order to, in a precise way, show how the
analytical solution for general $N$ develops. This will in particular, allow
us to see that, although the structure of the $R_{i}$ solutions is different
for odd and even $N$, the sum of $R_{i}$'s casts exactly in the same way.

\subsection{Three Cooper pairs}

We now consider the three Richardson's equations given in Eq. (26). We first
replace one of them by their sum which reads as Eq. (35) with three terms
instead of two. By keeping the $m=1$ term only in this sum, we find $%
z_{1}+z_{2}+z_{3}\simeq 0$. Since, in the one hand the $z_{i}-z_{j}$
differences have to be in $\sqrt{\gamma }$ for the Eqs. (26) to have a
solution with $z_{i}$ small, while in the other hand, $(z_{1},z_{2},z_{3})$
must have some internal symmetries, we are led to look for a solution which
reads at the lowest order in $\gamma $ as%
\begin{equation}
z_{1}=-z_{2}\simeq a_{1}\sqrt{\gamma },\text{ \ \ }z_{3}\simeq 0  \label{40}
\end{equation}%
When inserted into the first and third equations (26) in which the first
term only in each sum is kept, we find that these two equations reduce to
the same equation which reads as
\begin{equation}
0\simeq I_{1}a_{1}+\frac{1}{a_{1}}+\frac{1}{2a_{1}}  \label{41}
\end{equation}%
This readily gives $a_{1}^{2}=-3/2I_{1}$\ which again makes $a_{1}$
imaginary at the lowest order in $\gamma $. If we now keep two terms in the
sum of Eqs. (26), namely
\begin{equation}
0\simeq I_{1}(z_{1}+z_{2}+z_{3})+\frac{I_{2}}{2}\left(
z_{1}^{2}+z_{2}^{2}+z_{3}^{2}\right)  \label{42}
\end{equation}%
we find the sum of $z_{i}$'s at the lowest order in $\gamma $\ as%
\begin{equation}
z_{1}+z_{2}+z_{3}\simeq \frac{3}{2}\gamma \frac{I_{2}}{I_{1}^{2}}  \label{43}
\end{equation}%
When used into the energy of three pairs $E_{3}=R_{1}+R_{2}+R_{3}$, we find,
since $\epsilon _{c}\gamma =4/\rho _{0}$, this energy as
\begin{equation}
E_{3}=3\left( E_{1}+\frac{2}{\rho _{0}}\frac{I_{2}}{I_{1}^{2}}\right)
\label{44}
\end{equation}%
When compared to Eq. (38), this shows that the energy change per pair is
just twice as large for three pairs than for two pairs. Let us now see how
this nicely simple result extends when the number of pairs keeps increasing.

\subsection{Four and five Cooper pairs}

(i) In the case of four pairs, we have four equations similar to Eq. (25).
We replace one of them by their sum. Its lowest order term gives $%
z_{1}+z_{2}+z_{3}+z_{4}\simeq 0$. Since the $z_{i}-z_{j}$ differences must
be of the order of $\sqrt{\gamma }$ for the $z_{i}$'s to be small, the
expected symmetrical solution must read as%
\begin{equation}
z_{1}=-z_{4}\simeq a_{1}\sqrt{\gamma },\text{ \ \ }z_{2}=-z_{3}\simeq a_{2}%
\sqrt{\gamma }  \label{45}
\end{equation}%
When inserted into the three other equations, we find that two of them are
identical; so that we are left with two equations to get $a_{1}$ and $a_{2}$%
, namely%
\begin{eqnarray}
0 &\simeq &I_{1}a_{1}+\frac{1}{a_{1}-a_{2}}+\frac{1}{a_{1}+a_{2}}+\frac{1}{%
2a_{1}}  \notag \\
&\simeq &\left( I_{1}a_{1}^{2}+\frac{2a_{1}^{2}}{a_{1}^{2}-a_{2}^{2}}+\frac{1%
}{2}\right) \frac{1}{a_{1}}  \label{46}
\end{eqnarray}%
This imposes the prefactor of $1/a_{1}$ to be zero, with a similar result
for $1/a_{2}$. If we now add these two prefactors, we get
\begin{equation}
0\simeq I_{1}\left( a_{1}^{2}+a_{2}^{2}\right) +2+1  \label{47}
\end{equation}%
This gives $a_{1}^{2}+a_{2}^{2}\simeq -3/I_{1}$. When inserted into the sum
of Richardson's equations, with now two terms kept in the sum, we find $%
z_{1}+z_{2}+z_{3}+z_{4}\simeq 3\gamma I_{2}/I_{1}^{2}$. Since $\epsilon
_{c}\gamma =4/\rho _{0}$, the energy of four pairs ultimatly appears as%
\begin{equation}
E_{4}=4\left( E_{1}+\frac{3}{\rho _{0}}\frac{I_{2}}{I_{1}^{2}}\right)
\label{48}
\end{equation}

(ii) For five pairs, the sum of the five Richardson's equations taken at
lowest order in $\gamma $, gives $z_{1}+z_{2}+z_{3}+z_{4}+z_{5}\simeq 0$.
Since the differences $z_{i}-z_{j}$ must be in $\sqrt{\gamma }$ for the $%
z_{i}$'s to be small, we expect the symmetrical solution to read as
\begin{equation}
z_{1}=-z_{5}\simeq a_{1}\sqrt{\gamma },\text{ }z_{2}=-z_{4}\simeq a_{2}\sqrt{%
\gamma },\text{ }z_{3}\simeq 0  \label{49}
\end{equation}%
The remaining four equations are then identical two by two, so that we are
again left with two equations only for $a_{1}$ and $a_{2}$. They read%
\begin{eqnarray}
0 &\simeq &I_{1}a_{1}+\frac{1}{a_{1}-a_{2}}+\frac{1}{a_{1}}+\frac{1}{%
a_{1}+a_{2}}+\frac{1}{2a_{1}}  \notag \\
&=&\left( I_{1}a_{1}^{2}+\frac{2a_{1}^{2}}{a_{1}^{2}-a_{2}^{2}}+\frac{3}{2}%
\right) \frac{1}{a_{1}}  \label{50}
\end{eqnarray}%
and similar one for $a_{2}$. By adding the $1/a_{1}$\ and $1/a_{2}$
prefactors in the above equation, we get%
\begin{equation}
0\simeq I_{1}\left( a_{1}^{2}+a_{2}^{2}\right) +2+3  \label{51}
\end{equation}%
instead of Eq. (47). So that we now have $a_{1}^{2}+a_{2}^{2}=-5/I_{1}$.
When inserted into the sum of Richardson's equations in which we now keep
two terms in the sum, we find $z_{1}+z_{2}+z_{3}+z_{4}+z_{5}\simeq 5\gamma
I_{2}/I_{1}^{2}$; so that the energy of five pairs appears as%
\begin{equation}
E_{5}=5\left( E_{1}+\frac{4}{\rho _{0}}\frac{I_{2}}{I_{1}^{2}}\right)
\label{52}
\end{equation}

Although the numerical prefactors in the $z_{i}$\ sums do not seem to have
any physically relevant meaning, the results given in Eqs. (38), (44), (48),
and (52) for the energy of $N$ pairs clearly indicate that the energy change
\textit{per pair} depends linearly on the number of pairs as $(N-1)$, in
spite of the fact that the precise structure of the solution of Richardson's
equation is definitly different for $N=2n$ and $N=2n+1$. In order to better
grasp how this nice analytical result shows up, let us briefly consider one
more set of $N$'s, namely $N=6$ and 7, since these correspond to $n=3$ which
usually is generic while $n=2$ in the case of $N=(4,5)$ is usually not, 2
and 2! being equal.

\subsection{Six and seven Cooper pairs}

(i) For six Cooper pairs, the symmetrical solution is%
\begin{equation}
z_{1}=-z_{6}\simeq a_{1}\sqrt{\gamma },\text{ }z_{2}=-z_{5}\simeq a_{2}\sqrt{%
\gamma },\text{ }z_{3}=-z_{4}\simeq a_{3}\sqrt{\gamma }  \label{53}
\end{equation}%
Among the remaining five Richardson's equations, we only get three different
ones for $a_{1}$, $a_{2}$, $a_{3}$. They read%
\begin{eqnarray}
0 &\simeq &I_{1}a_{1}+\frac{1}{a_{1}-a_{2}}+\frac{1}{a_{1}-a_{3}}  \notag \\
&&+\frac{1}{a_{1}+a_{3}}+\frac{1}{a_{1}+a_{2}}+\frac{1}{2a_{1}}  \notag \\
&\simeq &\left( I_{1}a_{1}^{2}+\frac{2a_{1}^{2}}{a_{1}^{2}-a_{2}^{2}}+\frac{%
2a_{1}^{2}}{a_{1}^{2}-a_{3}^{2}}+\frac{1}{2}\right) \frac{1}{a_{1}}
\label{54}
\end{eqnarray}%
and similarly for $a_{2}$ and $a_{3}$. By adding the $1/a_{i}$ prefactors,
we end with%
\begin{equation}
0\simeq I_{1}\left( a_{1}^{2}+a_{2}^{2}+a_{3}^{2}\right) +2+2+2+3\frac{1}{2}
\label{55}
\end{equation}%
This leads to $z_{1}+...+$ $z_{6}\simeq \left( 15/2\right) \gamma
I_{2}/I_{1}^{2}$; so that since $\epsilon _{c}\gamma =4/\rho _{0}$, we get
the expected result, namely%
\begin{equation}
E_{6}\simeq 6\left( E_{1}+\frac{5}{\rho _{0}}\frac{I_{2}}{I_{1}^{2}}\right)
\label{56}
\end{equation}

(ii) For seven Cooper pairs, the symmetrical solution is
\begin{equation}
z_{1}=-z_{7}\simeq a_{1}\sqrt{\gamma },\text{ }z_{2}=-z_{6}\simeq a_{2}\sqrt{%
\gamma },\text{ }z_{3}=-z_{5}\simeq a_{3}\sqrt{\gamma }  \notag
\end{equation}
\begin{equation}
z_{4}\simeq 0  \label{57}
\end{equation}%
When inserted into the five remaining Richardson's equations, we only get
three different ones. They read as%
\begin{equation}
0\simeq I_{1}a_{1}+\frac{1}{a_{1}-a_{2}}+\frac{1}{a_{1}-a_{3}}+\frac{1}{a_{1}%
}+\frac{1}{a_{1}+a_{3}}+\frac{1}{a_{1}+a_{2}}+\frac{1}{2a_{1}}  \label{58}
\end{equation}%
The sum $a_{1}^{2}+a_{2}^{2}+a_{3}^{2}$ ends by reading as Eq. (55) except
for the last $1/2$ factor which is now replaced by $3/2$. This leads to $%
z_{1}+...+$ $z_{7}\simeq \left( 21/2\right) \gamma I_{2}/I_{1}^{2}$. So that
we find the result we expect, namely%
\begin{equation}
E_{7}\simeq 7\left( E_{1}+\frac{6}{\rho _{0}}\frac{I_{2}}{I_{1}^{2}}\right)
\label{59}
\end{equation}

\subsection{$2n$ and $2n+1$ Cooper pairs}

The above analysis of Richardson's equations for $N=1,2,3,...,7$ has made us
knowledgeable enough to now face the general $N$ solution through a similar
procedure.

(i) When the number of pairs is even, $N=2n$, we expect the symmetrical
solution for the $z_{i}$'s at the lowest order in $\gamma $ to be
\begin{equation}
z_{1}=-z_{2n}\simeq a_{1}\sqrt{\gamma },\text{ }z_{2}=-z_{2n-1}\simeq a_{2}%
\sqrt{\gamma }\text{, ..., }  \notag
\end{equation}
\begin{equation}
z_{n}=-z_{n+1}\simeq a_{n}\sqrt{\gamma }  \label{60}
\end{equation}%
By inserting these $z_{i}$'s into Richardson's equations, we get $n$
different equations for $a_{1}$, ..., $a_{n}$. They read as%
\begin{equation}
0\simeq I_{1}a_{1}+\frac{1}{a_{1}-a_{2}}+\frac{1}{a_{1}-a_{3}}+...  \notag
\end{equation}
\begin{equation}
+\frac{1}{a_{1}+a_{3}}+\frac{1}{a_{1}+a_{2}}+\frac{1}{2a_{1}}  \label{61}
\end{equation}%
When multiplied by $a_{1}$ and added to similar ones for $a_{2}$, ..., $%
a_{n} $, these equations lead to%
\begin{equation}
0\simeq I_{1}\left( a_{1}^{2}+...+a_{n}^{2}\right) +2C_{n}^{2}+\frac{n}{2}
\label{62}
\end{equation}%
where $C_{n}^{2}=n(n-1)/2$ is the number of ways to choose a couple ($%
a_{i}-a_{j}$) among $n$ coefficients $a_{i}$. Note that the above equation
agrees with Eq. (47) for $n=2$ and with Eq. (55) for $n=3$. When inserted
into the sum of Richardson's equations with now two terms kept in the sum,
namely%
\begin{equation}
0\simeq I_{1}\sum_{i=1}^{2n}z_{i}+\frac{I_{2}}{2}\sum_{i=1}^{2n}z_{i}^{2}
\label{63}
\end{equation}%
we find from Eqs. (60, 62) that, at lowest order in $\gamma $, the sum of $%
z_{i}$ is given by%
\begin{equation}
\sum_{i=1}^{2n}z_{i}\simeq \left[ 2n(n-1)/2+n/2\right] \gamma
I_{2}/I_{1}^{2}\simeq n(n-1/2)\gamma \frac{I_{2}}{I_{1}^{2}}  \label{64}
\end{equation}%
So that the sum of energy-like quantities introduced by Richardson appears
to read as%
\begin{eqnarray}
E_{2n} &=&R_{1}+...+R_{2n}=2nE_{1}+(z_{1}+...+z_{2n})\epsilon _{c}  \notag \\
&\simeq &2n(E_{1}+\frac{2n-1}{\rho _{0}}\frac{I_{2}}{I_{1}^{2}})  \label{65}
\end{eqnarray}%
This result casts into the structure we guessed for the energy change of $2n$
pairs as induced by the Pauli exclusion principle between an even number $%
N=2n$ of similar pairs.

(ii) If the number of pairs is odd, $N=2n+1$, we do the same but one of the $%
z_{i}$ is now equal to zero at the lowest order in $\sqrt{\gamma }$. Eq.
(61) is then replaced by%
\begin{eqnarray}
0 &\simeq &I_{1}a_{1}+\frac{1}{a_{1}-a_{2}}+\frac{1}{a_{1}-a_{3}}+...+\frac{1%
}{a_{1}-a_{n}}  \notag \\
&&+\frac{1}{a_{1}}+\frac{1}{a_{1}+a_{n}}+...+\frac{1}{a_{1}+a_{2}}+\frac{1}{%
2a_{1}}  \label{66}
\end{eqnarray}

So that the sum of $a_{n}$'s just reads as Eq. (62) with $n$/2 replaced by $%
3n/2$. Consequently, we now find, instead of Eq. (64)%
\begin{equation}
\sum_{i=1}^{n}z_{i}\simeq \left[ 2n(n-1)/2+3n/2\right] \gamma
I_{2}/I_{1}^{2}=n(n+\frac{1}{2})\gamma \frac{I_{2}}{I_{1}^{2}}  \label{67}
\end{equation}%
which ultimately leads to the expected result%
\begin{equation}
E_{2n+1}=R_{1}+...+R_{2n+1}  \notag
\end{equation}
\begin{equation}
\simeq \left( 2n+1\right) \left( E_{1}+\frac{2n}{\rho _{0}}\frac{I_{2}}{%
I_{1}^{2}}\right)  \label{68}
\end{equation}

\subsection{Validity of the result}

Let us now come back to the validity of the procedure leading to Eqs. (65)
and (68).

(i) We first wish to note that these equations allow us to cast the energy
of $N$ Cooper pairs into the same general form given in Eq. (1), for odd and
even $N$, the interaction energy $E_{int}$ in Eq. (1) being such that

\begin{eqnarray}
2E_{int} &=&\frac{1}{\rho _{0}}\frac{I_{2}}{I_{1}^{2}}\simeq \frac{1}{\rho
_{0}}(1+2y_{c})  \notag \\
&\simeq &\frac{1}{\rho _{0}}+\frac{\epsilon _{c}}{N_{\Omega }}=\frac{1}{\rho
_{0}}+\epsilon _{V}  \label{69}
\end{eqnarray}%
$1/\rho _{0}$ is the energy increase when adding one electron of each spin
at the frozen Fermi sea level when the density of states is constant and
equal to $\rho _{0}$. The number of pair states within the potential
extension $\Omega $ above $\left\vert F_{0}\right\rangle $ is $N_{\Omega
}=\rho _{0}\Omega $. The one-Cooper pair binding energy is $\epsilon _{c}$
while $\epsilon _{V}$ is the elementary contribution to the Cooper pair
binding energy from each up and down spin empty state feeling the attractive
potential. The result we find leads us to write the energy of $N$ electron
pairs added to $\left\vert F_{0}\right\rangle $\ as%
\begin{eqnarray}
E_{N} &\simeq &NE_{1}+\frac{N(N-1)}{2}E_{int}  \notag \\
&\simeq &N\left[ 2\left( \varepsilon _{F_{0}}+\frac{N-1}{2\rho _{0}}\right)
-\epsilon _{c}\left( 1-\frac{N-1}{N_{\Omega }}\right) \right]   \label{70}
\end{eqnarray}

The $N(N-1)/2$ prefactor in front of $E_{int}$\ clearly indicates that this
part of the energy comes from interaction between \textit{two} Cooper pairs.
Due to the very simplified form of the reduced BCS potential $\mathcal{V}$,
interaction \textit{between} Cooper pairs can only come from the Pauli
exclusion principle between electrons. However, since the Pauli exclusion
principle is $N$-body by essence, higher order terms in $N(N-1)(N-2)...$
should appear in the $N$-pair energy. A possible reason for their absence is
the fact that we perform a $\gamma $ expansion to analytically solve
Richardson's equations but we only keep the lowest order term. This leads us
to think that the result given in Eq. (70) is valid provided that
corrections induced by interactions are small compared to the main term,
namely $E_{1}=2\varepsilon _{F_{0}}-\epsilon _{c}$. Since there are two
physically different "main" terms, let us consider them separately.

(i) As already highly remarkable, the first term of this $N$-pair energy, $2%
\left[ N\varepsilon _{F_{0}}+N(N-1)/2\rho _{0}\right] $, is the \textit{exact%
} kinetic energy of $N$ free electrons with up and down spins when added at
the Fermi level $\varepsilon _{F_{0}}$, \textit{whatever} $N$ is - under the
assumption of a constant density of states $\rho _{0}$. Indeed, electrons
are fermions; the Pauli exclusion principle forces their energies to
increase as $\varepsilon _{F_{0}}$, $\varepsilon _{F_{0}}+1/\rho _{0}$, ...,
$\varepsilon _{F_{0}}+(N-1)/\rho _{0}$; so that the total energy of $N$ up
and down spin electrons is equal to
\begin{equation}
E_{N}^{(normal)}=2\sum_{p=0}^{N-1}\left( \varepsilon _{F_{0}}+\frac{p}{\rho
_{0}}\right) =2\left[ N\varepsilon _{F_{0}}+\frac{N\left( N-1\right) }{2\rho
_{0}}\right]   \label{71}
\end{equation}%
which exactly is the first term of Eq. (70). This shows that the procedure
we have used to get the $N$-pair energy, with a validity a priori restricted
to the dilute limit on the one-Cooper pair scale, still produces the exact
free part of this energy, whatever $N$. Since this free part is nothing but
the "normal" part of the superconductor energy, i.e., the one which does not
cancel for $V=0$, we can rewrite Eq. (70) as
\begin{equation}
E_{N}-E_{N}^{(normal)}\simeq -N\left( 1-\frac{N-1}{N_{\Omega }}\right)
\epsilon _{c}  \notag
\end{equation}%
\begin{equation}
\simeq -2N\epsilon _{V}[N_{\Omega }-(N-1)]  \label{72}
\end{equation}%
where $E_{N}^{(normal)}$\ is the energy of these $N$ pairs in a normal
state, i.e., without interaction, the $N$ added electrons having transformed
the "frozen" Fermi sea $\left\vert F_{0}\right\rangle $ into the "normal"
Fermi sea $\left\vert F\right\rangle $.

(ii) If we now turn to the correlated part of the $N$-pair energy, i.e., the
second term of Eq.(70) or the RHS of Eq.(72), a correction small compared to
the main term imposes $N\ll N_{\Omega }$, i.e., $N$ small compared to the
total number of empty states from which the one-Cooper pair bound state is
constructed. We note that this small $N$ scale is not at all related to the
one-Cooper pair scale which led to $N\ll N_{c}$.

(iii) From a mathematical point of view, the validity of Eq. (70) must be
traced back to the validity of the $\gamma $ expansion we have performed in
Eq.(27) and that we have used, within one or two terms only, this $\gamma $
expansion being a priori valid for $z_{i}$ small compared to 1. Due to Eqs.
(63, 64), the sum of the $z_{i}$ real parts as well as the sum of the
squares of the imaginary parts, increase as $N^{2}\gamma $. Since $\gamma $
scales as $1/N_{c}$, as seen from Eq. (34), the fact that we use an
expansion which is valid provided that all the $\left\vert z_{i}\right\vert $%
's are small, leads us to a priori restrict the validity of the obtained
result to $N$ small on the $N_{c}$ scale, or a power law of this scale,
condition which is far more restrictive than $N\ll N_{\Omega }$.
Consequently, we do not a priori expect the result we have obtained in Eq.
(70) to be valid in the dense BCS regime in which the number of added pairs
is half the total number of pair states $N_{\Omega }$ in the potential layer
$\Omega $.

(iv) Nevertheless, the fact that the expansion we have performed, gives the
correct normal part for the energy of $N$ pairs, even when $N\gg N_{c}$, is
a strong indication that our result is most probably valid on a wider range.
Another strong support of this idea is the fact that this result, when
extended to a complete filling of the potential layer, i.e., to $N=N_{\Omega
}$, gives an average pair binding energy which reduces to zero. This is
fully reasonable since all the available states are then occupied, so that
nothing can change when the attractive potential is turned on since, due to
its peculiar form, no exchange between pairs can exist. Consequently, the
energy difference between correlated and normal states must reduce to zero
when $N=N_{\Omega }$ in agreement with Eq. (72).

All this leads us to guess that, in spite of the mathematical procedure we
have used to get the expression of $E_{N}$ given in Eq. (70), which, to be
valid, a priori restricts the number of pairs to $N\ll N_{c}$, the validity
range of our result is most probably far larger that this quite restrictive
condition which corresponds to an extremely dilute limit, the scale being
the one-Cooper pair extension.

Let us now compare the $N$-pair energy we have obtained within this
canonical ensemble approach, with the standard BCS result obtained in the
grand canonical ensemble.

\section{Link with the BCS dense limit}

\subsection{The standard BCS result}

(i) One important characteristic feature of the BCS theory in its
traditional form is that there is only one energy scale associated to
pairing: the energy difference $\Delta $ between the ground state and the
first excited state linked to "pair-breaking" processes. Nevertheless, the
gap $\Delta $ is often used to reach some physical understanding for the
ground state \textit{itself}. The energy difference between the ground state
and the normal state having the same number of noninteracting electrons is
written in textbooks as%
\begin{equation}
U_{s}-U_{n}=-\frac{1}{2}\rho _{0}\Delta ^{2}  \label{73}
\end{equation}%
where $\Delta $ reads in terms of the potential extension and amplitude as $%
\Delta =2\omega _{c}e^{-1/\rho _{0}V}=\Omega e^{-1/\rho _{0}V}$\ \cite%
{extension}.

By comparing the gap with the single pair binding energy given in
Eq.(9), we note a factor of 2 difference in the prefactors, but also
in the exponent, the latter being crucial in the weak-coupling limit
because it makes the gap far larger than the single pair binding
energy. This immediately leads us to
reconsider the widely accepted connection between "Cooper pair energy" and $%
\Delta $ because this binding energy can hardly increase with pair number up
to the gap value due to Pauli blocking among the available states to form
paired states.

(ii) The usual way to understand the energy difference in Eq. (73) is to say
that each pair which lies in a narrow shell $\Delta $ at the noninteracting
electron Fermi level, has an energy of the order of $\Delta $. The number of
such pairs

\begin{equation}
N_{\Delta }\sim \rho _{0}\Delta   \label{74}
\end{equation}%
is very small compared to the total number of pairs feeling the potential.
This gives an energy difference between superconducting and normal states of
the order of $N_{\Delta }{\Delta }=\rho _{0}\Delta ^{2}$. This understanding
indeed reproduces the result of Eq. (73), but within a numerical $1/2$
prefactor only, which is not that easy to bring in.
%However, correlated pairs with an energy of the order of the gap have to be quitedifferent from the Cooper pairs we have considered in the Richardson'sprocedure, since their binding energy, said to be of the order of the gap $%\Delta $, essentially is $e^{1/\rho _{0}V}$ larger than the single Cooperpair binding energy $\epsilon _{c}$.

(iii) By noting that the energy difference between superconducting and
normal states also reads%
\begin{eqnarray}
U_{s}-U_{n} &=&-\frac{1}{2}\rho _{0}\Omega ^{2}e^{-2/\rho _{0}V}  \notag \\
&=&-\left[ \rho _{0}\frac{\Omega }{2}\right] \left[ \Omega e^{-2/\rho _{0}V}%
\right]   \label{75}
\end{eqnarray}%
there is another way to understand this result. Indeed the BCS configuration
corresponds to add $\rho _{0}\Omega /2=N_{\Omega }/2$ pairs to the frozen
Fermi sea $\left\vert F_{0}\right\rangle $, in order to fill half of the
potential layer (see Fig. 1b). Consequently, in order to recover the
standard BCS result given in Eq. (75), the average binding energy of these $%
N_{\Omega }/2$ pairs\ must be taken equal to $\Omega e^{-2/\rho _{0}V}$.
This quantity which is just half the binding energy of a single Cooper pair $%
\epsilon _{c}$, exactly corresponds to the average pair binding energy we
have found in Eq. (72), when $N$ is equal to $N_{\Omega }/2$. Since the
normal state energy $U_{n}$ is nothing but $E_{N}^{(normal)}$ appearing in
Eq. (70), we see that the BCS result given in Eq. (75) is identical to the
energy of $N$ Cooper pairs we have obtained in Eq. (72). Although the
validity of our result seemed at first to be mathematically reduced to the
very dilute limit $N\ll N_{c}$, due to the expansion (31) we have used to
get it and the fact that we have only kept the first order term of this
expansion, it is quite remarkable to note that its extension to the dense
regime gives the BCS result \textit{exactly}, i.e., with all its numerical
prefactors.

\subsection{Dense limit with a non-symmetrical potential}

Such an agreement is even more remarkable when we consider a potential
configuration which is not symmetrical with respect to the normal Fermi
level, i.e., for an arbitrary filling of the potential layer, not
nessesarily $1/2$ as in the usual BCS configuration.

(i) Following Eq. (2.33a) in Tinkham's textbook \cite{Tinkham}, the energy
gap obeys the equation%
\begin{eqnarray}
1 &=&\frac{\rho _{0}V}{2}\int_{-\varepsilon _{F}+\varepsilon
_{F_{0}}}^{\Omega -\varepsilon _{F}+\varepsilon _{F_{0}}}\frac{d\xi }{\sqrt{%
\xi ^{2}+\Delta ^{2}}}  \notag \\
&=&\frac{\rho _{0}V}{2}\left[ \sinh ^{-1}\left( \frac{\varepsilon
_{F_{0}}+\Omega -\varepsilon _{F}}{\Delta }\right) \right.   \notag \\
&&\left. +\sinh ^{-1}\left( \frac{\varepsilon _{F}-\varepsilon _{F_{0}}}{%
\Delta }\right) \right]   \label{76}
\end{eqnarray}%
In the symmetrical configuration, $\varepsilon _{F}$ is equal to $%
\varepsilon _{F_{0}}+\Omega /2$, so that the two terms in the above equation
are equal. The value of $\Delta $ in the weak coupling limit ($\rho _{0}V\ll
1$) is then readily obtained as%
\begin{equation}
\Delta \simeq \Omega e^{-1/\rho _{0}V}  \label{77}
\end{equation}

In a nonsymmetrical configuration, Eq. (76) cannot be solved exactly.
However, if the Fermi level $\varepsilon _{F}$ is located deeply inside the
layer, in order to have both $\varepsilon _{F_{0}}+\Omega -\varepsilon _{F}$
and $\varepsilon _{F}-\varepsilon _{F_{0}}$ much larger than $\Delta $, the
inverse hyperbolic sines can be replaced by logarithms. Under these
assumptions, Eq. (76) can be rewritten as%
\begin{equation}
1\simeq \frac{\rho _{0}V}{2}\ln \left[ 4\frac{\left( \varepsilon
_{F_{0}}+\Omega -\varepsilon _{F}\right) \left( \varepsilon _{F}-\varepsilon
_{F_{0}}\right) }{\Delta ^{2}}\right]   \label{78}
\end{equation}%
which then gives%
\begin{equation}
\Delta \simeq 2\left[ \left( \varepsilon _{F_{0}}+\Omega -\varepsilon
_{F}\right) \left( \varepsilon _{F}-\varepsilon _{F_{0}}\right) \right]
^{1/2}e^{-1/\rho _{0}V}  \label{79}
\end{equation}%
The usual BCS expression (77) for the gap $\Delta $ readily follows from
this general result when the potential is symmetrical, i.e., for $%
\varepsilon _{F}=\varepsilon _{F_{0}}+\Omega /2$.

(ii) If we now use the expression of the gap given in Eq. (79) to calculate
the energy of $N$ pairs, we find%
\begin{eqnarray}
U_{s}-U_{n} &=&\frac{\rho _{0}}{2}\left[ \left( \varepsilon _{F}-\varepsilon
_{F_{0}}\right) ^{2}\right.   \notag \\
&&-\left( \varepsilon _{F}-\varepsilon _{F_{0}}\right) \sqrt{\left(
\varepsilon _{F}-\varepsilon _{F_{0}}\right) ^{2}+\Delta ^{2}}  \notag \\
&&+\left( \varepsilon _{F_{0}}+\Omega -\varepsilon _{F}\right) ^{2}-\left(
\varepsilon _{F_{0}}+\Omega -\varepsilon _{F}\right)   \notag \\
&&\left. \sqrt{\left( \varepsilon _{F_{0}}+\Omega -\varepsilon _{F}\right)
^{2}+\Delta ^{2}}\right]   \label{80}
\end{eqnarray}%
For $\varepsilon _{F_{0}}+\Omega -\varepsilon _{F}$ and $\varepsilon
_{F}-\varepsilon _{F_{0}}$ both much larger than $\Delta $, the Taylor
expansion of the RHS of Eq. (80) shows that this energy difference takes the
same form as the one for a symmetric potential given in Eq. (73), namely%
\begin{equation}
U_{s}-U_{n}\simeq -\frac{1}{2}\rho _{0}\Delta ^{2}  \label{81}
\end{equation}%
with $\Delta $ now given by Eq. (79) instead of Eq. (77). Substitution of
Eq. (79) into Eq. (81) then gives%
\begin{eqnarray}
&&U_{s}-U_{n}  \notag \\
&\simeq &-\left[ \rho _{0}\left( \varepsilon _{F}-\varepsilon
_{F_{0}}\right) \right] \left[ 2\Omega e^{-2/\rho _{0}V}\right]   \notag \\
&&\left[ 1-\frac{\varepsilon _{F}-\varepsilon _{F_{0}}}{\Omega }\right]
\label{82}
\end{eqnarray}%
which also reads as%
\begin{eqnarray}
&&U_{s}-U_{n}  \notag \\
&=&-N\epsilon _{c}\left( 1-\frac{N}{N_{\Omega }}\right)   \notag \\
&=&-2N\epsilon _{V}(N_{\Omega }-N)  \label{83}
\end{eqnarray}%
The first factor in the RHS of Eq. (82) is the number $N$ of pairs in the
potential layer. The second factor of Eq. (82) is the binding energy $%
\varepsilon _{c}$ of one isolated Cooper pair while the last factor, which
reduces to 1/2 for a symmetrical potential, in agreement with Eq. (75), just
corresponds to the "shrinkage" of the average binding energy resulting from
the Pauli exclusion principle. Eq.(82) thus shows that, in the
nonsymmetrical configuration too, it is quite natural to take as Cooper pair
number, the number of pairs in the potential layer, their average binding
energy being linearly reduced, due to the Pauli exclusion principle. Eq.
(83) then shows that this binding energy is proportional to the number of
\textit{empty} states $(N_{\Omega }-N)$ remaining in the potential layer.

We also wish to note that this BCS-like grand canonical treatment can be
extended to the dilute regime of pairs provided that their number still is
macroscopically large, i.e., their density is finite. In this case, one has
to supplement the gap equation by an equation for the chemical potential.
The resulting expression of $U_{s}-U_{n}$ again agrees with Eq. (82).
Conceptually, this BCS analysis is similar to the Eagles's model\cite{Eagles}
for the density-induced BEC-BCS crossover (see also Ref. \cite%
{Tonycross,Strinati,Engel}). Although the BEC-BCS crossover problem is
currently considered mostly within cold gases \cite{roland,reviews1}, there
are some indications that a density-induced crossover also has some
relevance for correlated pairs in high-$T_{c}$ cuprates \cite%
{Levin,Gantmakher}.

\subsection{Cooper pairs as composite bosons}

From a general point of view, the energy of $N$ interacting pairs can always
be written as%
\begin{equation}
NE_{1}+\frac{N(N-1)}{2}E_{int}(n)  \tag{84}  \label{84}
\end{equation}%
where $n=N/L^{D}$ is the pair density, $L^{D}$ being the sample volume. The
result of Eq. (70), obtained in the dilute limit, gives $E_{int}(n%
\rightarrow 0)=E_{int}$. Due to the reduced potential taken in the BCS
theory, with a constant interaction between electrons having opposite spin
and opposite momenta, it is hard to physically expect higher order density
contributions in the interaction term coming from the potential itself. A
possible $N$ dependence of $E_{int}(n)$ is more likely to come from the
Pauli exclusion principle which is $N$-body by essence. Actually, the BCS
result clearly indicates that this additionnal $N$ dependence does not
exist. It can be of interest to note that we have recently found a similar
result for the ground state energy of $N$ Frenkel excitons: terms
proportional to third and higher powers in $N$ are absent in the energy of $N
$ Frenkel excitons \cite{paper3}. The Hamiltonian for Frenkel excitons \cite%
{paper1} has similarities with the BCS Hamiltonian because it only couples
electrons to holes localized on the \textit{same} atomic sites: in the same
way, the interaction potential used in the BCS theory only couples up and
down spin electrons having \textit{opposite} momenta: in both cases, this is
a one-to-one coupling between different fermions, so that Cooper pairs like
Frenkel excitons are composite bosons with one degree of freedom only, in
contrast to Wannier excitons which have two: the electron momentum and the
hole momentum.

As a last comment, we wish to mention that the interaction term $E_{int}$
has two contributions which have completely different origins although both
of them end with the same $N(N-1)$ dependence. These two different origins
nicely enlighten the fact that Cooper pairs are boson-like particles made of
two fermions. Indeed, one part of $E_{int}$ is associated to a kinetic
energy change coming from electrons added to the frozen Fermi sea $%
\left\vert F_{0}\right\rangle $. The sum of their kinetic energy increases $%
1/\rho _{0}+2/\rho _{0}+...+(N-1)/\rho _{0}$, equal to $N(N-1)/2\rho _{0}$
for each spin, just gives the normal part of $E_{int}$. The $N(N-1)$
prefactor of this free part comes from a sum of $N$ \textit{different}
energies with a arithmetical progression. In contrast, the interaction
energy also has a $N(N-1)$ contribution which is fully linked to the bosonic
nature of the Cooper pairs, these pairs \textit{all} having the same binding
energy $2[N_{\Omega }-(N-1)]\varepsilon _{V}$ coming from the number of
empty states in the potential layer: the resulting energy change then is $%
2(N-1)\varepsilon _{V}=\varepsilon _{c}(N-1)/N_{\Omega }$. When multiplied
by $N$, this also leads to a $N(N-1)$ prefactor in the energy of $N$ pairs,
but this prefactor results from the sum of $N$ \textit{identical} energy
changes. This understanding nicely reveals the two aspects of Cooper pairs:
The bosonic nature of the pairs shows up in the correlated part of $E_{int}$%
, while their fermionic nature appears in the normal part of the energy.
This makes Cooper pairs quite interesting composite bosons, definitely,
worth to be tackled along the ideas we have recently developed in the
composite boson many-body theory\cite{Monique}, originally constructed for
Wannier excitons.

\section{Conclusions}

We have extended the well-known Cooper's model beyond one pair and revealed
the deep link which exists between this model and BCS superconductivity.
While Cooper has considered one electron pair added to a frozen Fermi sea,
in a region where a small attracting potential acts, we here consider a
large number of pairs, this number being increased one by one, starting from
$N=1$. We calculate the ground state energy of these $N$ pairs resulting
from both, the standard BCS potential and the Pauli exclusion principle.
This is done by solving Richardson's equations for $N$ energy-like
quantities, their sum giving the exact $N$-pair energy. We determine these $N
$ quantities analytically through an expansion in the dimensionless
parameter associated to the inverse of sample volume which turns out to also
be the inverse of the number of pairs from which single Cooper pairs would
start to overlap. By keeping the first order term in this dilute limit, we
find that the energy of $N$ pairs reads as the energy of $N$ individual
Cooper pairs, within a $N(N-1)$ correction which makes the average pair
energy varying \textit{linearly} with pair number. The \textit{\ average
pair binding energy is found to be proportional to the number of empty
states feeling the potential}. Although this result only uses the first
order term of an expansion which is mathematically valid in the dilute limit
on the one-Cooper pair scale, quite remarkably, this linear behavior stays
valid up to the dense BCS regime, even when the attractive potential between
electrons is not symmetric with respect to the normal Fermi level, or when
the added pairs completely fill the layer where the potential acts. The
physics of this interaction term can be traced back to the Pauli exclusion
principle, the number of available empty states within the potential
extension decreasing linearly when the pair number increases. Since this
argument is not affected when going from the dilute limit to the dense limit
on the one-Cooper pair scale, it may appear as physically reasonable to find
that the energy change obtained for $N$ small on this scale, extends to the
dense BCS limit in which a large fraction of the states feeling the
potential is occupied.

The nontrivial result we here obtain, brings some new light on the concept
of "Cooper pair number" in the standard BCS result for condensation energy.
The usual understanding is based on the idea that the Cooper pair binding
energy is of the order of the excitation gap $\Delta $; these pairs are
concentrated in a very thin shell around the normal Fermi level, its width
of the order of $\Delta $ being much smaller than the width of the layer
where the attractive potential between up and down spin electrons acts.
However, in this understanding, the "number of Cooper pairs" is more a
qualitative concept than a rigorously defined quantity. The understanding we
here propose uses the very basic fact that correlations exist between
\textit{all }the electrons with opposite spins and opposite momenta lying in
the potential layer. The binding energy we find for such pairs in the $N$%
-pair configuration is much smaller than $\Delta $, being half the energy of
one isolated pair in the case of a BCS potential extending symmetrically on
both sides of the normal Fermi level. This binding energy reduction entirely
comes from the Pauli exclusion principle which reduces the number of states
available to form bound pairs. This makes the average pair binding energy
ultimately proportional to the number of empty states. Such a link between
an isolated Cooper pair and the $N $-pair configuration is going to be of
importance for considering superconductivity in the framework of the
many-body theory for composite bosons we have recently developed, Cooper
pairs definitely being quite interesting composite bosons with
characteristics rather different from the ones of Wannier excitons which
have up to now been the main application of this new many-body theory \cite%
{Monique}.

\acknowledgments

One of us (M. C.) wishes to thank Tony Leggett for enlightening discussions
during her invitations at the University of Illinois in Urbana. The other
(W. V. P.) acknowledges supports from the French Ministry of Education
during his stay in Paris, the Dynasty Foundation, and the Russian Foundation
for Basic Research (project no. 09-02-00248). Both of us have also
beneficiated from useful discussions with Roland Combescot and Christiane
Caroli at the begining of this work.


\begin{references}

\bibitem{Frol}H. Frohlich, Phys. Rev. {\bf 79}, 845 (1950).

\bibitem{Cooper}L. N. Cooper, Phys. Rev. {\bf 104}, 1189 (1956).

\bibitem{BCS}J. Bardeen, L. N. Cooper, and J. R. Schrieffer, Phys. Rev. {\bf 108}, 1175
(1957).

\bibitem{Bogoliubov}N. N. Bogoliubov, Usp. Fiz. Nauk \textbf{67}, 549 (1959).

\bibitem{Delft} F. Braun and J. von Delft, Phys. Rev. Lett. \textbf{81}, 4712 (1998).

\bibitem{Rich1}R. W. Richardson, Phys. Lett. {\bf 3}, 277 (1963).

\bibitem{Rich2}R. W. Richardson and N. Sherman, Nucl. Phys. {\bf 52}, 221
(1964).

\bibitem{Rich3}R. W. Richardson, J. Math. Phys. {\bf 18}, 1802
(1977).

\bibitem{Bogoliubov1}For the applicability of the canonical ensemble to
the theory of superconductivity, see N. N. Bogoliubov, Usp. Fiz. Nauk {\bf 67}, 549 (1959)
[Sov. Phys. Usp. {\bf 2}, 236 (1959)].

\bibitem{Gaudin}M. Gaudin, J. Phys. (Paris) {\bf 37}, 1087
(1976).

\bibitem{review}J. Dukelsky, S. Pittel, and G. Sierra, Rev. Mod.
Phys. {\bf 76}, 643 (2004).

\bibitem{pap3}W. V. Pogosov, M. Combescot, and M. Crouzeix, Phys. Rev.
B {\bf 81}, 174514 (2010).

\bibitem{pap1}W. V. Pogosov and M. Combescot.
Pis'ma v ZhETF {\bf 92}, 534  (2010) [JETP Letters {\bf 92}, 534  (2010)].

\bibitem{Fetter}A. L. Fetter and J. D. Walecka,
{\it Quantum Theory of Many-Particle Systems}, Dover Publications, New York
(2003).

\bibitem{Tinkham}M. Tinkham, {\it Introduction to
Superconductivity}, Dover Publications, New York (2004).

\bibitem{Leggett}A. J. Leggett, {\it Quantum liquids: Bose condensation
and Cooper pairing in condensed-matter systems}, Oxford University Press (2006).

\bibitem{Schrieffer}J. R. Schrieffer,
{\it Theory of Superconductivity}, Perseus Books Group, Massachusetts
(1999).

\bibitem{Roman}J. M. Roman, G. Sierra, and J. Dukelsky,
Nucl. Phys. B {\bf 634}, 483 (2002).

\bibitem{Altshuler}E. Yuzbashyan, A. A. Baytin, and B. L. Altshuler,
Phys. Rev. B {\bf 71}, 094505 (2005).

\bibitem{Monique}M. Combescot, O. Betbeder-Matibet, and F. Dubin,
Physics Reports {\bf 463}, 215 (2008) and references therein.

\bibitem{extension}Note that, in most textbooks, the one-Cooper pair
problem is discussed with an interaction potential extending between $%
\varepsilon _{F_{0}}$\ and $\varepsilon _{F_{0}}+\omega _{c}$ (see for
example, Eq. (2.6) in Tinkham's book \cite{Tinkham}), so that $\omega _{c}$
must be identified with our $\Omega $, the binding energy of one Cooper pair
then reading as $\epsilon _{c}\simeq 2\omega _{c}e^{-2/\rho _{0}V}=2\Omega
e^{-2/\rho _{0}V}$. When turning to the BCS
dense regime, the same textbooks consider a potential extending
symmetrically on both sides of the Fermi level between $\varepsilon
_{F}-\omega _{c}$ and $\varepsilon _{F}+\omega _{c}$, so that the
potential extension then reads as $\Omega =2\omega _{c}$. This notation change
 a priori irrelevant, however masks the importance of the total number of empty
states available to make the Cooper-paired states and the associated deep
connection which exists between the one-Cooper pair problem and the BCS
regime. To our opinion, a very important part of superconductor physics lies in
this empty state number: it must be preserved through coherent notations.


\bibitem{Eagles}D. M. Eagles, Phys. Rev. {\bf 186}, 456 (1969).

\bibitem{Tonycross} A. J. Leggett, J. de Physique. Colloques {\bf 41}, C7 (1980);
A. J. Leggett, {\it Proceedings of the XVIth Karpacz Winter School of
  Theoretical Physics, Karpacz, Poland},  pp. 13-27, Springer-Verlag (1980).

\bibitem{Strinati}N. Andrenacci, A. Perali, P. Pieri, and G. C. Strinati,
Phys. Rev. B {\bf 60}, 12410 (1999).

\bibitem{Engel} C. A. R. S\ddag\ de Melo, M. Randeria, and J. R.
Engelbrecht, Phys. Rev. Lett. \textbf{71}, 3202 (1993); M. Randeria, J.-M.
Duan, and L.-Y. Shieh, Phys. Rev. Lett. \textbf{62}, 981 (1989).

\bibitem{roland} R. Combescot, X. Leyronas, and M. Y. Kagan, Phys. Rev. A
\textbf{73}, 023618 (2006).

\bibitem{reviews1} I. Bloch, J. Dalibard, and W. Zwerger, Rev. Mod. Phys.
\textbf{80}, 885 (2008); S. Giorgini, L. P. Pitaevskii, and S. Stringari,
Rev. Mod. Phys. \textbf{80}, 1215 (2008).

\bibitem{Levin} Q. Chen, J. Stajic, S. Tan, and K. Levin, Physics Reports
\textbf{412}, 1 (2005).

\bibitem{Gantmakher} V. F. Gantmakher and V. T. Dolgopolov, Usp. Fiz. Nauk
\textbf{180}, 3 (2010) [Phys. Usp. \textbf{53}, 1 (2010)].

\bibitem{paper3}W. V. Pogosov and M. Combescot, Eur. Phys. J. B {\bf 68}, 183 (2009);

\bibitem{paper1}M. Combescot and W. V. Pogosov, Phys. Rev. B {\bf 77}, 085206 (2008).


\end{references}
\end{document}